# Significant enhancement of compositional and superconducting homogeneity in Ti rather than Ta-doped Nb$_3$Sn

C. Tarantini,[1,a] Z.H. Sung,[1] P.J. Lee,[1] A.K. Ghosh,[2] and D.C. Larbalestier[1]

[1]*National High Magnetic Field Laboratory, Florida State University, Tallahassee, FL 32310, USA*

[2]*Magnet Division, Brookhaven National Laboratory, Upton, NY 11973, USA*

Nb$_3$Sn wires are now very close to their final optimization but despite its classical nature, detailed understanding of the role of Ta and Ti doping in the A15 is not fully understood. Long thought to be essentially equivalent in their influence on $H_{c2}$, they were interchangeably applied. Here we show that Ti produces significantly more homogeneous chemical and superconducting properties. Despite Ta-doped samples having a slightly higher $T_c$ onset in zero-field, they always have a wider $T_c$-distribution. In particular, whereas the Ta-doped A15 has a $T_c$-distribution extending from 18 down to 5-6 K (the lowest expected $T_c$ for the binary A15 phase), the Ti-doped samples have no A15 phase with $T_c$ below ~12 K. The much narrower $T_c$ distribution in the Ti-doped samples has a positive effect on their in-field $T_c$-distribution too, leading to an extrapolated $\mu_0H_{c2}(0)$ 2 Tesla larger than the Ta-doped one. Ti-doping also appears to be very homogeneous even when the Sn content is reduced in order to inhibit breakdown of the diffusion barriers in very high $J_c$ conductors. The enhanced homogeneity of the Ti-doped samples appears to result from its assistance of rapid diffusion of Sn into the filaments and by its incorporation into the A15 phase interchangeably with Sn on the Sn sites of the A15 phase.

[a] Electronic mail: tarantini@asc.magnet.fsu.edu

Application of large quantities of Nb$_3$Sn conductors first for ITER[1] and now for the High Luminosity upgrade of the Large Hadron Collider (LHC)[2] has motivated major recent R&D and raised multiple questions about how best to optimize the superconducting strand,[3] especially with respect to balancing the conflicting requirements of small hysteretic loss, high critical current density ($J_c$) and a high residual resistance ratio (RRR) in the Cu stabilizer.[4] However, few recent studies have investigated the fundamental properties of the A15 phase in these newer wires, which has a composition range with $T_c$ varying from 18 K (stoichiometric) to 6 K on the Sn-poor (18at.%Sn) side of the binary. Controlling this composition range is crucial to optimizing Nb$_3$Sn properties. The classical nature of Nb$_3$Sn implies broadly well-known properties.[5,6] For example the very similar effect of Ti and Ta in enhancing $H_{c2}$ was clearly shown many years ago by Suenaga *et al.*,[7] who also noticed a slight increase of $T_c$ upon doping that was greater for Ta. Ta, Ti or Ti+Ta doping are now standard for enhancing the in-field performance.[8] However, for a long time Ti and Ta were both thought to substitute on the Nb site, but more recent studies suggested that Ti can substitute on the Sn site.[9,10] A comparison of the in-field physical and compositional properties of Ti and Ta-doped A15 phase in the most modern, highest-$J_c$ wires taking into account these factors has not been performed yet. Maintaining a high RRR in the Cu stabilizer is crucial but frequently in conflict with the need to obtain the largest possible quantity of high quality (i.e. homogeneous A15 close to the stoichiometric composition to give it the highest $T_c$) A15 phase. Finding a compromise heat treatment (HT) allows sufficient Sn diffusion for the formation of A15 phase with high Sn content without jeopardizing the Nb (or Nb alloy) diffusion barrier (DB) integrity while avoiding the grain growth that degrades the pinning properties.[11] An alternative approach is to reduce the Sn fraction (i.e. increase the Nb:Sn ratio) in the wire, as recently introduced into their design by Oxford Superconducting Technology (OST),[12] to avoid reacting through the DB but potentially sacrificing A15 quality. Here we will investigate 1) subtle but important differences between Ti and Ta doping that show that they are not equivalent and 2) an important insensitivity to Nb:Sn ratio that can greatly benefit RRR while avoiding a significant penalty in A15 properties.

In this paper we compare the properties of three recent small sub-element high-$J_c$ Nb$_{1-x}$Sn$_x$ Restacked-Rod-Process (RRP®) wires manufactured by OST: two wires are Ti-doped but with different Nb:Sn ratios (3.4:1 and 3.6:1 for Ti-doped Standard-Sn "Ti-stdSn" and for Ti-doped Reduced-Sn "Ti-redSn" samples, respectively, including the DB)[12], whereas the third wire is Ta-doped with standard Sn amount ("Ta-stdSn"). All samples have the same wire architecture (108/127 stack design, 0.778 mm wire diameter, ~50 μm sub-element diameter, 0.2 Cu:Nb local area ratio). Each RRP® sub-element is made of Nb rods dispersed in a Cu matrix surrounding a Sn-rich core and assembled in a Cu-clad Nb DB.[13] In the case of Ta-doping, Nb-7.5wt.%Ta (4 at.%Ta) replaces Nb in both the DB and the rods. In the Ti-doped case, some Nb-47 wt.%Ti (Nb-Ti) rods are evenly distributed in the rod pack. The Sn content is changed by varying the diameter of the Sn-core in the sub-element center: in Ti-redSn the Sn-core is ~5% smaller by volume than in the standard design.[12] During the initial 210°C/48h+400°C/48h HT, Sn diffuses into the filament pack (with Cu counter-diffusing into the core),[14] whereas the final A15 phase-formation high-temperature HT is performed at 662-665°C/48h. The principal properties are summarized in TABLE I. Despite its reduced Sn-content, Ti-redSn has a $J_c$ only 3.6% lower than Ti-stdSn (2930 versus 3035 A/mm$^2$ at 12 T) and an identical irreversibility field at 4.2 K as determined by Kramer extrapolation ($\mu_0 H_K$ ~ 25.5 T). The most important difference is in the residual resistivity ratio (RRR): the performed HT is too aggressive for Ti-stdSn, producing a much suppressed RRR (36) due to Sn diffusion through the DB, whereas Ti-redSn has an optimal RRR of 181. Ta-stdSn has a $J_c$ slightly larger than Ti-redSn (2950 A/mm$^2$), RRR still larger than 100 but its irreversibility field is significantly reduced ($\mu_0 H_K$ ~ 23.8 T). To understand what induces these differences we performed microscopic analysis in a Zeiss 1540 Crossbeam® Field Emission Scanning Electron Microscope (FESEM) using metallographically polished transverse cross-sections. Energy Dispersive Spectroscopy (EDS) was performed at 15 kV using standard-less analysis (EDAX TEAM V3-4) with an EDAX Apollo XP SDD detector. To determine the intrinsic superconducting properties, specific heat characterization of the wires was performed up to ~15 T in a Quantum Design 16 T physical property measurement system (PPMS).

TABLE I Physical and material properties of the studied RRP® wires

|  | Ti-stdSn[a] | Ti-redSn | Ta-stdSn[a] |
|---|---|---|---|
| Billet | 14895FE | 14896FE | 12879 |
| Doping | Ti | Ti | Ta |
| Sn content | Standard Sn | 5% less Sn | Standard Sn |
| Heat treatment | 662°C/48h | 662°C/48h | 665°C/48h |
| $J_c$(12T), A/mm$^2$ | 3035 | 2927 | 2952 |
| $\mu_0 H_K$, T | 25.45 | 25.47 | 23.84 |
| RRR | 36 | 181 | 106 |

[a] Not optimized heat treatment.

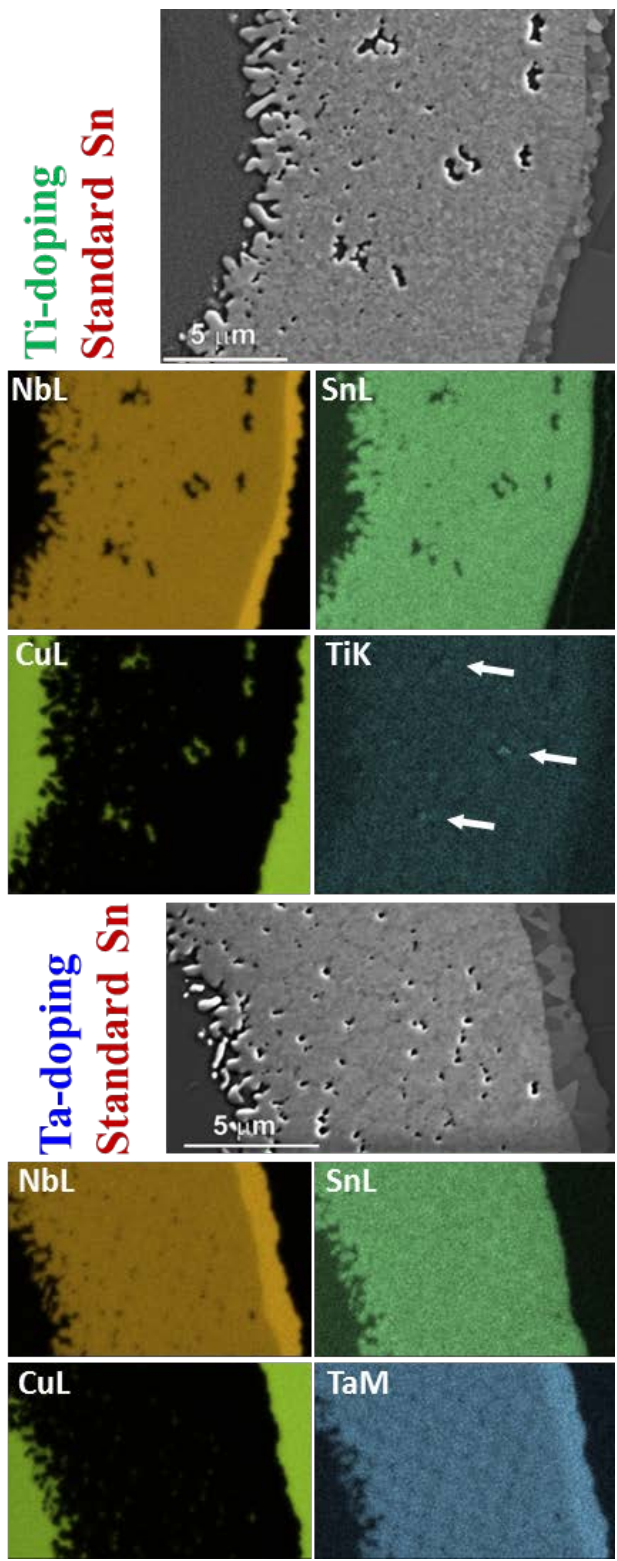

**Figure 1** FESEM-BSE images and qualitative (brighter=higher atomic composition) maps [Nb(L), Sn(L), Cu(L), Ti(K) or Ta(M)] obtained by FESEM–EDS for Ti-stdSn (top 5 images) and Ta-stdSn (lower 5 images) RRP® wires. The white arrows indicate high Ti regions in the Ti-stdSn wire.

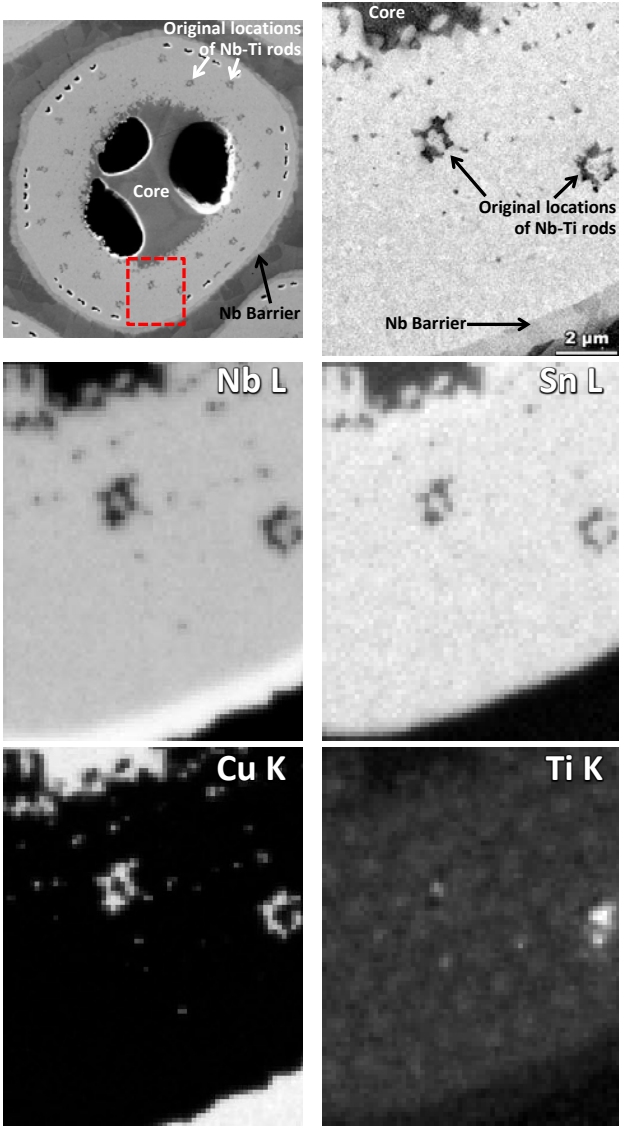

**Figure 2** FESEM-BSE images (top 2 images) and qualitative (brighter=higher atomic composition) maps [Nb(L), Sn(L), Cu(L), Ti(K)] obtained by FESEM–EDS for Ti-redSn RRP® wire (lower 4 images).

Fig.1 compares the SEM images and EDS maps of Ti-stdSn and Ta-stdSn across the A15 layer. Ti-stdSn appears remarkably uniform with no obvious variation in the Sn composition across the layer. The most notable features are some residual Cu regions and a few high Ti regions at the original Nb-Ti filament locations. Ta-stdSn has noticeably more Sn and Ta compositional fluctuations: the Ta-map shows a clear honeycomb structure corresponding to the original filament structure all across the A15 layer and this is also seen, but to a less marked extent, in the Sn-map. Apart from the DB thickness, no obvious difference was found between the two Ti-wires. At higher magnification (Fig.2) the chemical maps for Ti-redSn confirm the Sn uniformity and reveal that the residual Ti is mainly present in the Ti sources furthest from the core, whereas the inner Ti sources are significantly Ti-depleted with Cu counter-diffusing in its place. Interestingly a high Ti content was detected at each filament center, normally the most Sn-poor A15 regions.

In order to quantify the A15 layer Sn gradient we performed EDS measurements of local Sn at the approximate centers of each filament. Fig. 3(a) shows that the Sn content varies from ~25% near the core to 22-22.5% near the DB without significant sample-to-sample difference. Considering the pseudobinary A15 phase and taking into account the different doping sites for Ti and Ta ($Nb_{1-x}(Sn_{1-y}Ti_y)_x$ and $(Nb_{1-y}Ta_y)_{1-x}Sn_x$),[9] we found that the Nb/ Sn+Ti ratio for Ti-stdSn and Ti-redSn varies from ~2.8 to ~3.2 with a wide central region of the layer close to the stoichiometric value, 3 (Fig.3(b)). In

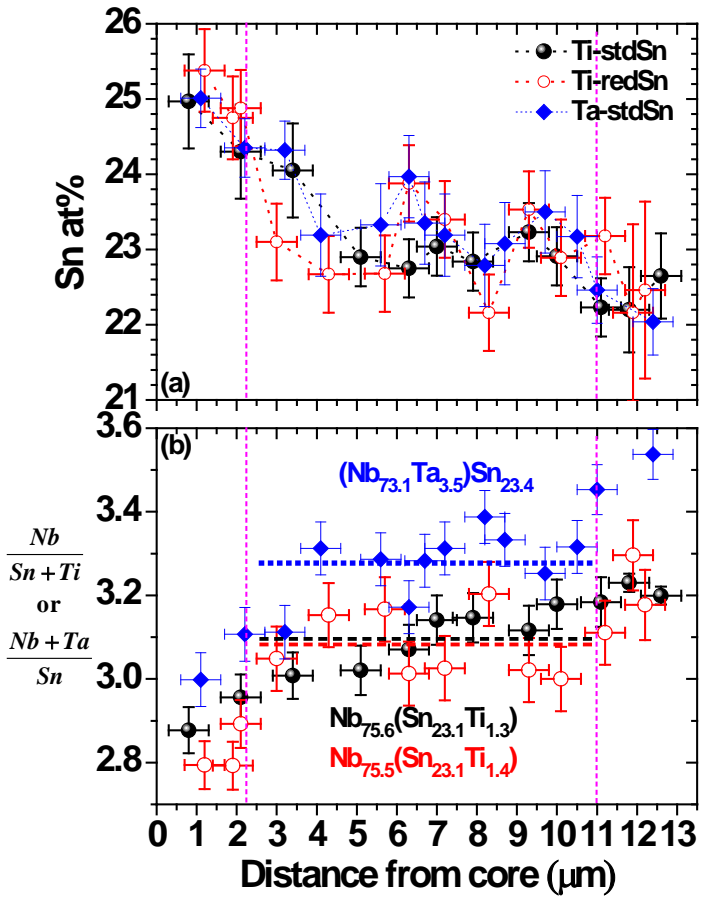

**Figure 3** (a) Sn atomic composition and (b) Nb/ Sn+Ti or Nb+Ta/Sn ratios across the A15 layer for tree different RRP® wires. The typical EDS sampled area is ~1μm-diameter as represented by the error bars. The compositions are calculated averaging on the central regions of the A15 layers.

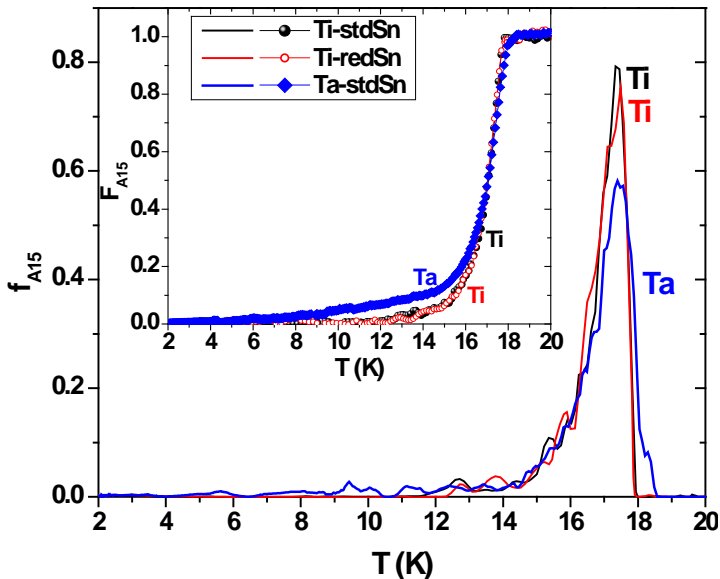

**Figure 4** $T_c$-distribution of the A15 phase, $f_{A15}$, and its integral, $F_{A15}$, (in the inset) for tree different RRP® wires.

contrast, for Ta-stdSn the Nb+Ta/Sn ratio changes from ~3 near the core to ~3.5 at the DB with the central region distinctly Sn-poor at ~3.3. In the center of the A15-layer the average Nb is 75.5-75.6at.% for the Ti-doped strands and 73.1% for Ta-stdSn, consistent with Ti sitting on Sn site.

The $T_c$-distribution determined by specific heat allows an independent estimate of the phase homogeneity. The $T_c$-distribution of the A15 phase alone, $f_{A15}$, and its integral, $F_{A15}$, were determined by the two-phase model developed in ref. 15 following an approach similar to that described in ref. 16. This model *a priori* excludes the Nb contribution, allowing the estimation of the A15 $T_c$-distribution above and also below the Nb transition. Fig.4 reveals that, despite Ta-stdSn having a $T_c$ onset ~0.5 K higher than the Ti-doped samples, its transition is broader, generating suppressed peak intensity in the $T_c$-distribution. Remarkably there are no significant differences between Ti-stdSn and Ti-redSn. In the inset, the integral $F_{A15}$ clearly shows that Ta-stdSn has A15 phase with $T_c$ as low as 5.5-6 K, whereas neither Ti-doped samples have A15 phase with $T_c$ below ~12 K.

The in-field $T_c$-distributions were studied up to ~15 T by analyzing the specific heat curves as done for the zero-field data.[15] For clarity the in-field $T_c$- distributions reported in Fig.5 are vertically shifted proportionally to the applied field. Fig.5(a) shows that in-field too there are no significant differences between Ti-stdSn and Ti-redSn. However, comparing Ti and Ta doping [Fig.5(b)] reveals that the Ta-stdSn peak intensity is smaller at every field and that the higher $T_c$ onset advantage of Ta at zero-field rapidly disappears in increasing field. For instance, at ~15 T, the majority of the $T_c$-distribution for Ta-stdSn is about 0.4 K lower in temperature than for Ti-stdSn. The Ta/Ti differences are clearly visible in the in-field $F_{A15}$ curves (e.g. upper inset of Figure 6). From these we estimate that Ti-stdSn and Ti-redSn still have about 50-51% of the A15 phase in the superconducting state at ~15 T, whereas in Ta-stdSn it is only 42%.

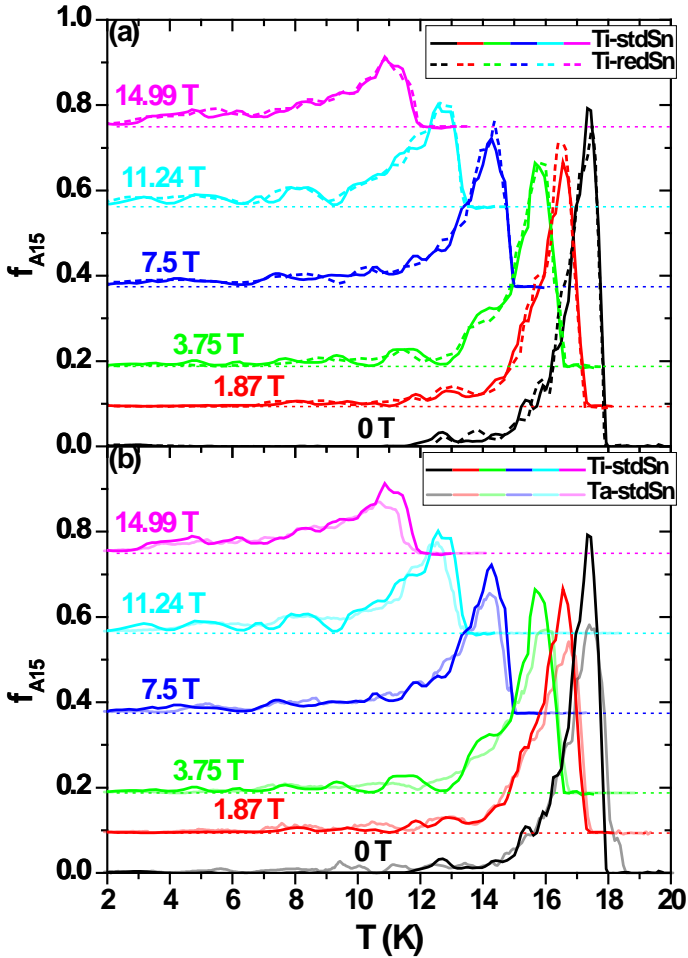

**Figure 5** In field $T_c$-distribution of the A15 phase, $f_{A15}$, for (a) Ti-stdSn and Ti-Red-Sn and for (b) Ti-stdSn and Ta-stdSn RRP® wires. The curves are vertically shifted proportionally to the applied field.

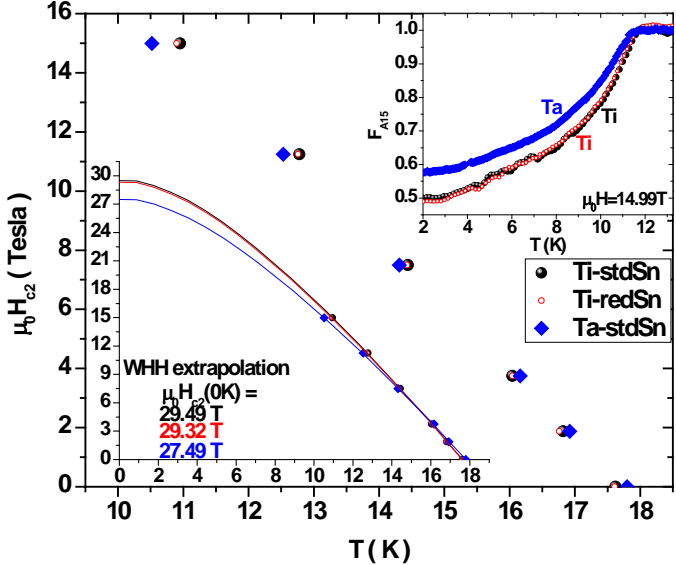

**Figure 6** Temperature dependence of $H_{c2}$ as determined by a 90% criterion on the $F_{A15}$ curves for three different RRP® wires. In the lower inset a WHH extrapolation to 0 K is performed. In the upper inset the $F_{A15}$(~15T) curves are plotted.

To better estimate the differences in the in-field performance we calculated the temperature dependence of $H_{c2}$ from the 90% criterion of the $F_{A15}$ curves (Fig.6). The Ti-doped samples show essentially identical $H_{c2}$ curves, whereas Ta-stdSn has an obviously lower $H_{c2}(T)$. WHH extrapolation[17] to 0 K determines that the Ti-doped samples reach a $\mu_0H_{c2}$ of 29.3-29.5 T, whereas Ta-stdSn has $\mu_0H_{c2}$ ~ 27.5 T, about 2 Tesla lower. The $\mu_0H_{c2}$ crossover between Ti and Ta doping occurs at about ~5.0-5.5 T.

In considering the broader implications of these results, we note first that the micro-chemical analysis

(Fig.1) reveals that the Ti-doped wires produce a significantly more uniform A15 phase than does Ta-doping, in which a filament-scale honeycomb inhomogeneity is visible in both Sn and Ta maps. This kind of high-Sn, low-Ta ring structure was already observed in the larger sub-element size 54/61 Ta-doped RRP® wires,[11] suggesting that it is intrinsic to Ta-doped strands: either Ta slows Sn diffusion into the filament centers or alternatively Ti promotes Sn diffusion producing the more homogenous distribution of Sn into the Ti-doped filaments. Interesting also is the high Ti content at the center of each filament (Fig.2) despite the Ti sources being sparsely distributed in the filament pack: this denotes a high Ti-diffusivity that strongly enhances the A15 phase homogeneity and encourages a Nb/Sn+Ti ratio close to 3. Comparing the Nb/Sn+Ti and Nb+Ta/Sn ratios (Fig.3b) Ta-stdSn has a ratio of 3.3 in the central region of the layer implying a significant Sn deficiency and degraded superconducting performance.

There are distinct differences in the breadth of the $T_c$-distribution in Ti and Ta-doped wires (Fig.4) demonstrating how wide the range of composition is even in high-performance Ta-doped wires. The non-equilibrium nature of the A15 reaction is made clear by the observation of A15 $T_c$-distribution down to ~ 6 K, which is the lowest expected $T_c$ for Sn-poor A15 (<18.5%Sn).[5,10] In contrast, neither of the Ti-doped wires have a superconducting phase with $T_c$ below ~12 K, indicating that Ti significantly improves the phase homogeneity with respect to Ta. Comparing the Ta results obtained here with the ones previously measured on the 54/61 Ta-doped RRP® wires heat treated over a wide temperature range[11] (the data reported there were reanalyzed with the two-phase model), we found that the similarly heat treated 54/61 (665°C/50h) has a broad transition and superconducting phase with $T_c$ down to ~6 K comparable to the present 108/127 Ta-stdSn. Moreover we verified that, in order to obtain a flat $T_c$-distribution below 12 K that is similar to the Ti-doped samples, the Ta-doped 54/61 has to be heat treated at 750 °C for 96 hours, a temperature which is impractical for applications because of the substantial RRR degradation and reduced vortex pinning produced by significant A15 grain growth.

High magnetic fields accentuate even small property changes. The similar $T_c$-distributions at every field for Ti-stdSn and Ti-redSn [Fig.5(a)] prove their substantially identical A15 phase homogeneity, despite their different Nb:Sn ratios. In contrast the stronger in-field degradation of the $T_c$-distribution for Ta-stdSn [Fig.5(b)] implies that Ta-doped A15 is more field dependent than the Ti-doped wire. The suppressed $H_K$ estimated in Ta-stdSn could arise from either inferior intrinsic (low $H_{c2}$) or extrinsic material properties (low density of pinning centers). However the 2 T smaller $H_{c2}(0K)$ extrapolation determined by specific heat for Ta-stdSn establishes that its inferior field performance is mainly related to its lower $H_{c2}$.

In conclusion, we show that reduced Sn content in RRP® wires, introduced to mitigate RRR degradation,[12] surprisingly does not affect A15 phase quality nor substantially change its in-field performance, making this variant preferable for magnet applications. We have also found that Ti-doped wires have substantially more homogeneous $T_c$-distribution, even in fields up to 15 T. The origin of this greater homogeneity appears to be rapid diffusion of Ti during the wire HT and its capability to substitute on the Sn site. When the A15 layer compositions in the Ti and Ta-doped wires are assessed by their Nb/Sn+Ti and Nb+Ta/Sn ratios, the Ti-doped layers are much closer to stoichiometry, consistent with their much better $T_c$-distributions. Although all three wires have $J_c$(12T,4.2K)~2980±50 A/mm$^2$, the 2 T larger $H_{c2}$ of the Ti-doped wires means that they will be favored for higher field use.

**Acknowledgments**

RRP®® strand was provided under the LARP (LHC-Accelerator Research Program). This work was supported by the US Department of Energy (DOE) Office of High Energy Physics under award numbers DE-FG02-07ER41451 and DE-SC0012083, by the National High Magnetic Field Laboratory (which is supported by the National Science Foundation under NSF/DMR-1157490) and by the State of Florida. Work at BNL is supported by the US Department of Energy under Contract No. DE-AC02-98CH10886.

References

[1] A. Devred, I. Backbier, D. Bessette, G. Bevillard, M. Gardner, C. Jong, F. Lillaz, N. Mitchell, G. Romano, and A. Vostner, Superconductor Science and Technology **27**, 044001 (2014).

[2] L. Bottura, G. de Rijk, L. Rossi, and E. Todesco, IEEE Transactions on Applied Superconductivity **22**, 4002008 (2012).

[3] A. Ghosh, E. A. Sperry, J. D'Ambra, and L. D. Cooley IEEE Trans. Appl. Supercond. **19**, 2580 (2009); T. Boutboul, L. Oberli, A. den Ouden, D. Pedrini, B. Seeber, and G. Volpini, IEEE Trans. Appl. Supercond., **19**, 2564 (2009).

[4] L. Bottura, and A. Godeke, Rev. Accl. Sci. Tech. 05, 25-50 (2012).

[5] H. Devantay, J. L. Jorda, M. Decroux, J. Muller, and R. Flükiger, J Mater Sci **16**, 2145 (1981).

[6] T. P. Orlando, E. J. McNiff, Jr., S. Foner, and M. R. Beasley, Phys. Rev. B **19**, 4545 (1979).

[7] M. Suenaga, D. O. Welch, R. L. Sabatini, O. F. Kammerer, and S. Okuda, J. of App. Phys. **59**, 840 (1986).

[8] J. A. Parrell, M. B. Field, Y. Zhang, and S. Hong, IEEE Trans. Appl.Supercond. **15**, 1200 (2005).

[9] R. Flükiger, C. Senatore, M. Cesaretti, F. Buta, D. Uglietti, and B. Seeber, Supercond.Sci. Technol. **21** 054015 (2008).

[10] R. Flükiger, D. Uglietti, C. Senatore, and F. Buta, Cryogenics **48** 293-3072008 (2008).

[11] C. Tarantini, P. J. Lee, N. Craig, A. Ghosh, and D. C. Larbalestier, Supercond.Sci. Technol. **27** 065013 (2014).

[12] M. B. Field, Y. Zhang, H. Miao, M. Gerace, and J. A. Parrell, IEEE Trans. Appl. Supercond. **24**, 6001105 (2014).

[13] J. A. Parrell, Y. Zhang, R. W. Hentges, M. B. Field, and S. Hong, IEEE Trans. Appl. Supercond. **13** 3470 (2003).

[14] I. Pong, L.R. Oberli, L. Bottura, Supercond. Sci. Technol. **26**, 105002 (2013).

[15] C. Tarantini, C. Segal, Z. H. Sung, P. J. Lee, L. Oberli, A. Ballarino, L. Bottura, and D. C. Larbalestier, Supercond. Sci. Technol. **28**, 095001 (2015).

[16] Y. Wang, C. Senatore, V. Abächerli, D. Uglietti, and R. Flükiger, Supercond. Sci. Technol. **19**, 263 (2006).

[17] N. R. Werthamer, E. Helfard, and P. C. Hohenberg, Phys. Rev. **147** 295(1966).